# An improved transform theory for estimation of number density distribution of colloidal particles on a surface: A method for colloidal-probe atomic force microscopy

Ken-ichi Amano[*] and Taira Ishihara

*Department of Energy and Hydrocarbon Chemistry, Graduate School of Engineering, Kyoto University, Kyoto 615-8510, Japan. [*]E-mail: amano.kenichi.8s@kyoto-u.ac.jp*

## Main text

In this short letter, we explain an improved transform theory for colloidal-probe atomic force microscopy (CP-AFM). CP-AFM can measure a force curve between the colloidal probe and a wall surface in a colloidal dispersion. The transform theory can estimate the normalized number density distribution of the colloidal particles on the wall ($g_{WP}$) from the force curve measured by CP-AFM.

CP-AFM has been used for studying the effects of pH,[1] ionic strength,[2] and polyelectrolyte solutions[3] on the force curve. When the colloidal suspension has a certain volume fraction, an oscillatory force curve is obtained, and hence it has also been used for study of stratification of the colloidal particles on the wall.[4,5,6] However, the oscillatory force curve does not directly indicate $g_{WP}$. To obtain $g_{WP}$ from CP-AFM, transformation from the force curve into $g_{WP}$ is required. Hence, the short letter explains the transform theory which is made based on statistical mechanics of a simple liquid. The transform theory proposed here is an improved version of our previous theory for solvation structure.[7]

The transform theory requires a particle-induced force curve ($f_P$) between the colloidal particle and the wall. $f_P$ is obtained by subtracting $f_{abs}$ from $f_{pre}$, where $f_{abs}$ and $f_{pre}$ represent the fitted force curves obtained in the absence and presence of the colloidal particles, respectively. That is, an equation $f_P = f_{pre} - f_{abs}$ is applied for obtention of the particle-induced force curve. In theory, it is hypothesized that the surface properties (materials) of the colloidal probe and the substrate are the same. Considering a monodisperse colloidal dispersion, a flat wall surface, and a spherical colloidal probe, $f_P$ can be expressed as,[7]

$$\frac{f_P(s)}{2\pi} = \int_{s-2r_{\text{eff}}-r_{\text{CP}}}^{s-r_{\text{eff}}} P(l)\,(s - r_{\text{eff}} - l)\,dl, \quad (1)$$

where $\pi$, $s$, and $r_{CP}$ represent the circular constant, the separation between the substrate surface and the center of the colloidal probe, and the radius of the colloidal probe, respectively. $l$ is vertical distance between the centers of the colloidal particles contacting at the substrate and the colloidal probe, and $r_{eff}$ is effective radius of the colloidal particle contacting on the colloidal probe. $P$ represents the pressure between two flat surfaces, where the surface properties are the same. Value of $2r_{eff}$ is estimated from the local minimum position in $\partial f_P/\partial s$ near the position of the fist peak in $f_P$. The final value of $2r_{eff}$ is determined self-consistently in the calculation. Generally, $2r_{eff}$ is greater than the core diameter ($d_P$) of the colloidal particle and is less than $(1/\rho_0)^{1/3}$. The value of $r_{eff}$ does not significantly affect the calculation result, because $r_{CP}$ is much larger than $r_{eff}$ in the CP-AFM system. To obtain $P$ from $f_P$, the local maximum position in $\partial f_P/\partial s$ near the fist peak position in $f_P$ is estimated, and the values of $f_P$ from the local maximum position to a sufficiently far position are substituted into $\boldsymbol{F}^*$ in Eq. (2):

$$\boldsymbol{F}^* = \boldsymbol{H}\boldsymbol{P}, \qquad (2)$$

where $\boldsymbol{F}^*$ corresponds to the left-hand side of Eq. (1), $\boldsymbol{P}$ and $\boldsymbol{H}$ correspond to $P(l)$ and the other parts of Eq. (1), respectively. $\boldsymbol{H}$ is a square matrix whose variables are $l$ and $s$. A detailed explanation of $\boldsymbol{H}$ has been provided in our previous paper.[7] $\boldsymbol{P}$ can be obtained numerically using the inverse matrix of $\boldsymbol{H}$. Subsequently, $\boldsymbol{P}$ is converted to $P(l)$. Next, a value of $g_{WP}$ at the first peak ($g_c$) is calculated by:

$$g_c = -P_{\min}/(k_B T \rho_0), \qquad (3)$$

where $k_B$, $T$, and $\rho_0$ are the Boltzmann constant, the absolute temperature, and the bulk number density of the colloidal particles, respectively. $P_{\min}$ is the minimum value of $P$. When reliability of $f_P$ near the wall is low, $g_c$ should be calculated by substituting the maximum value of $P$ ($P_{\max}$) into an equation below[8,9]

$$g_c = \frac{1}{2} + \frac{\sqrt{k_B^2 T^2 \rho_0^2 + 4 k_B T \rho_0 P_{\max}}}{2 k_B T \rho_0}, \qquad (4).$$

because a position $l_{\max}$ corresponding to $P_{\max}$ is comparatively far from the wall in comparison with a position $l_{\min}$ corresponding to $P_{\min}$. In Eq. (4), Kirkwood superposition approximation[10,11,12] is used. Finally, $g_{WP}$ is obtained as follows:

$$g_{WP}(d_P/2 + l - l_{\min}) = 0 \qquad \text{for} \quad l < l_{\min}, \qquad (5a)$$

$$g_{WP}(d_P/2 + l - l_{\min}) = \frac{P(l)}{k_B T \rho_0 g_c} + 1 \qquad \text{for} \quad l_{\min} \leq l \leq l_0, \qquad (5b)$$

$$g_{WP}(d_P/2 + l - l_{\min}) = \left(\frac{P(l)}{k_B T \rho_0 g_c^\alpha} + 1\right)^{\frac{1}{\alpha}} \qquad \text{for} \quad l_0 < l, \qquad (5c)$$

where $l_0$ corresponds to a position $P(l) = 0$ near $l_{\min}$. When Eq. (4) is used for calculation of $g_c$, $l_{\min}$ is determined so as to $g_c$s estimated from Eqs. (3) and (4) are the same. If a negative value is obtained in the parenthesis on the right-hand side of Eq. (5c) due to low accuracy of the input data, $g_{WP}(d_P/2 + l - l_{\min})$ is calculated as follows:[7,13] $\{\exp[P(l)/(k_B T \rho_0 g_c^\alpha)]\}^{(1/\alpha)}$ when $P(l) < 0$ and $[P(l)/(k_B T \rho_0 g_c^\alpha) + 1]^{(1/\alpha)}$ when $P(l) \geq 0$. If a position where the wall-wall steric repulsion becomes infinity ($l_{str}$) is known, the first peak position in $g_{WP}$ is moved in parallel to $(l_{\max} - l_{str})/2$.[8] This movement comes from an analogy from the wall-wall rigid body system. $\alpha$ is a parameter for modified Kirkwood superposition approximation:

$$g_{\text{WPW}}(x;h) = \left(g_{\text{WP}}(x)g_{\text{WP}}(h-x)\right)^{\alpha} \qquad (6)$$

where $g_{\text{WPW}}$ represents the normalized number density of the colloidal particle within the two flat walls. $h$ and $x$ are the surface separation between two flat walls and an arbitrary position (height) in the confined space, respectively. The value of $\alpha$ is calculated as follows:

$$\alpha = \ln\left(\frac{k_B T \rho_0 + \sqrt{(k_B T \rho_0)^2 + 4 k_B T \rho_0 P_{\max}}}{2 k_B T \rho_0}\right) / \ln(g_c). \qquad (7)$$

Using the transform theory, we estimated $g_{\text{WP}}$ from a force curve measured by Piech and Walz.[5] Then, we found from the transformation that the stratification of $g_{\text{WP}}$ is enhanced as the surface potentials of the colloidal particles are increased (not shown and this result has already been obtained in autumn 2016). To check the validity of the transformed result, we conducted a pure theoretical calculation in spring 2017. In the calculation, an integral equation theory (Ornstein-Zernike equation coupled with hypernetted-chain closure: OZ-HNC)[14,15] is used. In the calculation, small spherical particles with certain volumes are used as the colloidal particles and the solution is treated as the continuous fluid. For the pair potentials between colloidal particle-colloidal particle and between colloidal particle-substrate, DLVO with constant surface charge potentials are introduced.[16] The pair potential between the colloidal particles ($V_{\text{PP}}$) is given by

$$V_{\text{pp}}(H) = -\pi \varepsilon_0 \varepsilon_r d_P \psi_P^2 \ln\left(1 - \exp(-H/D)\right), \qquad (8)$$

where $\varepsilon_0$ and $\varepsilon_r$ are the electric permittivity of vacuum, the relative permittivity of water, respectively.[17] $H$, $D$, and $\psi_P$, respectively, are the surface-surface separation between the colloidal particles, Debye length, surface potential of the colloidal particle. The pair potential between the substrate and the colloidal particle ($V_{WP}$) is given by

$$V_{wp}(H) = \frac{\pi\varepsilon_0\varepsilon_r d_P(\psi_p^2 + \psi_w^2)}{2}\left\{\frac{2\psi_p\psi_w}{\psi_p^2 + \psi_w^2}\ln\left(\frac{1+\exp(-H/D)}{1-\exp(-H/D)}\right) - \ln\left(1 - \exp(-\frac{2H}{D})\right)\right\}. \quad (9)$$

In Eq. (9), $H$ and $\psi_W$ represent the surface-surface separation between the wall and the colloidal particle and surface potential of the wall, respectively. We calculated the radial distribution function between the colloid particles in the bulk using 'a radial symmetric OZ-HNC'. Substituting the radial distribution function into 'OZ-HNC for between a flat wall and spherical particles', we calculated $g_{WP}$. In all of the OZ-HNC calculations, the number of meshes is 16384, the grid spacing is $d_P/100$ nm, and $T$ is set at 298 K.

In Fig. 1, effect of change in $\psi_P$ on $g_{WP}$ is shown, where $\psi_P$ is changed from 0 to −150 mV. In it, the parameters are as follows: $\psi_W$ = −20 mV, $D$ = 8.8 nm, $d_P$ = 21 nm, $\rho_0$ = 4.7×10$^{21}$ particles/m$^3$, $\varphi$ (volume fraction) = 0.023, $T$ = 298 K, and the relative permittivity is calculated using an equation proposed by Malmberg and Maryott.[17] As |$\psi_P$| increased (from 0 mV to −150 mV), the repulsion between the substrate and colloidal particle is increased, which may cause decay of the stratification. However, the behavior shown in Fig. 1 is different from the prediction. The heights of the peaks in $g_{WP}$ are increased by an increase in |$\psi_P$|. Mechanism of the stratification is explained as follows: When |$\psi_P$| is high, the colloidal particles push each other in the bulk and the space is relatively crowded. In this case, some of the colloidal particles are driven to the wall surface despite of the repulsion between the substrate and colloidal particle. Then, the crowding in the bulk is somewhat relaxed by the stratification.

In summary, we have proposed the transform theory from the force curve measured by CP-AFM into the number density distribution of the colloidal particles on the wall surface. We have conducted the transform theory and the verification calculation (OZ-HNC). We have found that the stratification is enhanced with increasing the surface potentials of the colloidal particles. Although there are approximations in the transform theory, we believe that the transform theory enable us to qualitatively or semi-quantitatively estimate the layer structure. Then, the mechanism of the stratification will also be studied by using the transform theory.

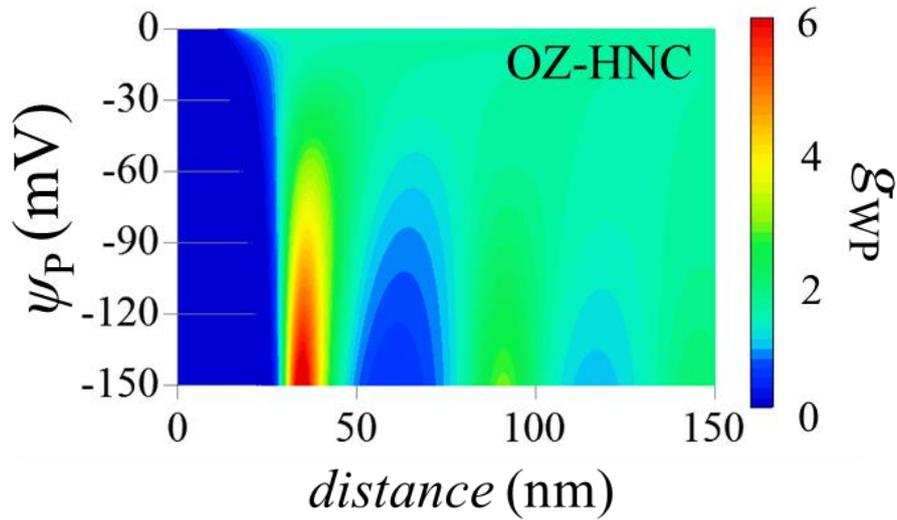

Fig. 1 Effect of change in $\psi_P$ on $g_{WP}$, where the horizontal axis represents the distance between the wall surface and the center of the colloidal particle.

## Acknowledgements

We thank Kota Hashimoto for supporting the preparation of Fig. 1. This work was supported by Grant-in-Aid for Young Scientists (B) from Japan Society for the Promotion of Science (15K21100).